\documentclass[12pt]{article}
\usepackage[a4paper, total={7in, 10in}]{geometry}
\usepackage[parfill]{parskip}
\usepackage{physics, tensor, float, subcaption}
\usepackage{graphicx}
\graphicspath{ {Plots/} }
\usepackage{jhep-mod}
\usepackage{bm}
\usepackage{soul}
\usepackage{amssymb,amsmath,amsthm}
\usepackage{mathrsfs}
\usepackage[utf8]{inputenc}
\usepackage{enumerate}
\usepackage{bigints}
\usepackage{xcolor}
\usepackage{appendix}
\usepackage{graphicx}
\usepackage{float}
\usepackage{tikz}
\usepackage{setspace}
\usepackage{cancel}
\usepackage{array}
\usepackage{tabulary}
\usepackage{doi}
\definecolor{purple}{rgb}{1,0,1}
\definecolor{lime}{HTML}{A6CE39} 

\definecolor{lime}{HTML}{A6CE39}
\newcommand{\orcidicon}{%
	\begin{tikzpicture}
	\draw[lime, fill=lime] (0,0) 
		circle [radius=0.16] 
		node[white] {{\fontfamily{qag}\selectfont \tiny ID}};
	\draw[white, fill=white] (-0.0625,0.095) 
		circle [radius=0.007];
	\end{tikzpicture}
	\hspace{-5mm}
}
\newcommand\orcidJosh{{\href{https://orcid.org/0000-0003-1200-7261}{\orcidicon}}}
\newcommand\orcidMatt{{\href{https://orcid.org/0000-0003-1088-6485}{\orcidicon}}}

\renewcommand{\O}{\mathcal{O}}

\newcommand{\be}{\begin{equation}gin{equation}}
\newcommand{\ee}{\end{equation}}

\def\rg{r_\gamma}
\def\ro{r_*}
\def\rp{r_\gamma}
\def\O{{\mathcal{O}}}
\def\sign{\mathrm{{sign}}}
\begin{document}
\newcommand{\arXiv}[1]{arXiv:\href{https://arxiv.org/abs/#1}{\color{blue}#1}}

\title{\huge{
{Perturbative photon escape cones\\  in the Kerr spacetime}
}}
\author{
\Large
Joshua Baines\!\orcidJosh,  {\sf  and} Matt Visser\!\orcidMatt}
\affiliation{School of Mathematics and Statistics, Victoria University of Wellington, 
\\
\null\qquad PO Box 600, Wellington 6140, New Zealand.}
\emailAdd{joshua.baines@sms.vuw.ac.nz}
\emailAdd{matt.visser@sms.vuw.ac.nz}
\renewcommand{\arXiv}[1]{arXiv:\href{https://arxiv.org/abs/#1}{\color{blue}#1}}
\def\L{{\mathcal{L}}}

\abstract{
\vspace{1em}

We consider the perturbative, fully explicit, analytical behaviour of photon escape cones in the Kerr spacetime. When one conducts the fully general non-perturbative Kerr analysis, one quickly finds that one must at some point appeal to numerical and/or graphical methods. Herein we find that we are able to say much more if we look at the slow rotation limit (i.e. $a\ll m$). Indeed we give explicit and tractable expressions for the first and second order (in $a$) contributions to both the shape of the escape cone, and the solid angle subtended by the escape cone. We then look at a few special cases at each order, thereby leading to explicit expressions for the black hole silhouette, expressions which are of great interest to the observational community when studying images of black hole silhouettes (``shadows'').

\bigskip
\noindent
{\sc Date:} Wednesday 20 March 2024; Thursday 4 April; Thursday 17 April;\\
\null\qquad\qquad   \LaTeX-ed \today


}

\maketitle
\def\tr{{\mathrm{tr}}}
\def\diag{{\mathrm{diag}}}
\def\cof{{\mathrm{cof}}}
\def\pdet{{\mathrm{pdet}}}
\def\QED{ {\hfill$\Box$\hspace{-25pt}  }}
\def\d{{\mathrm{d}}}
\def\sign{\hbox{sign}}

\parindent0pt
\parskip7pt

\clearpage
\null
\vspace{-75pt}
\section{Introduction}

A great deal of work has recently been done on calculations (and the generation of images) of the silhouettes, the ``shadows'', cast by black holes of various types~\cite{Cunha:2018, Gralla:2019,  Bambi:2019,  Abdujabbarov:2016, Hioki:2009, Johannsen:2010,  Broderick:2013, Broderick:2008, Johannsen:2015, Cardoso:2016, Claudel:2000, Tsukamoto:2014, Tsukamoto:2017, Ogasawara:2019mir, Ogasawara:2020frt, Zulianello:2020cmx}.
In part this surge of interest is due to the recent direct astrophysical observations of these silhouettes/shadows~\cite{EHT:2019a, EHT:2019d, EHT:2019e, EHT:2019f, EHT:2022a, EHT:2022e, EHT:2022c, EHT:2022f, Vagnozzi:2022}. In counterpoint, these silhouettes/shadows are just the time reversed complement of the escape cones of photons emitted from regions near the horizon, a topic which has its own long history (now almost 60 years). Synge started this line of work in the mid 1960s, showing that the opening angle $\Theta_0$ of the escape cone for photons emitted from a radius $r_*>r_H=2m$ in the Schwarzschild spacetime is given by~\cite{Synge:1966}
\begin{equation}
\sin\Theta_0=\sqrt{27}\;\frac{m}{r_*}\;\sqrt{1-\frac{2m}{r_*}}.
\end{equation}
Equivalently
\begin{equation}
\cos\Theta_0= - \left(1-\frac{3m}{r_*}\right)\sqrt{1+\frac{6m}{r_*}}.
\end{equation}
Hence, the solid angle subtended by the escape cone is given by
\begin{equation}
(\Delta\Omega_*)_0=\int_0^{2\pi} \int_0^{\Theta_0}\sin\vartheta\; \d\vartheta \; \d\varphi = 2\pi[1-\cos\Theta_0] 
= 2\pi\left[1+\left(1-\frac{3m}{r_*}\right)\sqrt{1+\frac{6m}{r_*}}\,\right].
\end{equation}
From this, one can calculate the ratio of photons which escape to future null infinity to the total number of photons emitted. Physically, this tells us the fraction of light which is able to escape from an isotropic light emitting object as it approaches the event horizon. This is a non-trivial physical effect which could, in theory, be observed. 
For future reference we note that for the Schwarzschild spacetime
\begin{equation}
\label{E:Sch-special}
\left.(\Delta\Omega_*)_0\right|_{r_*=2m} = 0; \qquad
\left.(\Delta\Omega_*)_0\right|_{r_*=3m} = 2\pi; \qquad
\left.(\Delta\Omega_*)_0\right|_{r_*=\infty} =4\pi.
\end{equation}
Recently, the present authors considered several explicit calculations of the escape cones, relevant to much more general static, spherically symmetric spacetimes~\cite{Baines:2023}. Given this, a natural question to then ask is this: Can this fully explicit discussion be generalised to stationary, axisymmetric \emph{rotating} spacetimes such as the Kerr spacetime? (For additional background, see references~\cite{Kerr1,Kerr2,Kerr3,Kerr4, Kerr5,Kerr6,Kerr7,Kerr8,Kerr9}.)  Herein we discuss what has been done on this subject, and focus specifically on what more can be said explicitly and analytically. 

\clearpage
\section{Preliminaries: Some exact results for Kerr}\enlargethispage{30pt}

Perlick and Tsupko give an extensive review on calculating black hole shadows in reference \cite{Perlick:2021}, specifically addressing the Kerr spacetime. Perlick and Tsupko give the two key equations determining the edge of the Kerr photon escape cone as
\begin{equation}\label{sintheta_general}
\sin\Theta =  {2\rg\sqrt{\rg^2-2m\rg+a^2} \sqrt{\ro^2-2m\ro+a^2} 
\over \ro^2(\rg-m) +\rp( \rp^2 -3 \rp m+2a^2)};
\end{equation} 
\begin{equation}\label{sinPhi_general}
\sin\Phi =- {\rg^2(\rg-3m)+ a^2[\rg  + m + (\rg-m)\sin^2\theta_* ]
\over 2a\rg \sin\theta_* \sqrt{\rg^2-2m\rg+a^2}}.
\end{equation}
Here the coordinates $(\theta_*,r_*)$ now denote the declination and radial position of the emission point, and the coordinates $(\Theta,\Phi)$ now denote the edge of the escape cone on the celestial sphere of the emission point.
The awkward part of the calculation is that $(\Theta,\Phi)$ are determined parametrically in terms of the (\emph{a priori} unknown) location $r_\gamma$ of the ``spherical photon orbits'', rapidly forcing one to adopt numerical techniques.\footnote{The ``spherical photon orbits'' are constant-$r$ orbits in the Boyer--Lindquist $r$ coordinate. They sweep out (a portion of) the constant $r$ topological 2-sphere. (This is not geometrically a constant-curvature 2-sphere). Except for one special case where the entire 2-sphere is swept out, the ``spherical photon orbits'' sweep out an equatorial zone of the topological 2-sphere.
}
See also references \cite{Tsupko:2017,Grenzebach:2014}. 
It is this need for a parametric representation leading to numerical analysis that we seek to avoid in the current article, aiming for as much as possible in the way of explicit formulae.

We can always re-write these parameteric equations for $(\Theta,\Phi)$ in the perhaps simpler but still exact form
\begin{equation}
\label{sintheta_general2} 
\sin\Theta =  {2\rg\sqrt{\rg^2-2m\rg+a^2} \sqrt{\ro^2-2m\ro+a^2} 
\over 2\rp( \rp^2 -2 m \rp+a^2) + (\ro^2-\rp^2)(\rp-m)};
\end{equation} 
\begin{equation}
\sin\Phi =- {\rg(\rg^2-2m\rg+a^2) - m(\rg^2-a^2) + a^2(\rg-m)\sin^2\theta_* 
\over 2a\rg \sin\theta_* \sqrt{\rg^2-2m\rg+a^2}}.
\end{equation}
Here the light is assumed to be emitted isotropically from the point $(r_*,\theta_*)$ and the angles $(\Theta,\Phi)$ describe the edges of the escape cone in suitable angular coordinates on the emitter's celestial sphere. 

Note that equations \eqref{sintheta_general} or \eqref{sintheta_general2}  can be recast into an equation for $\cos\Theta$ as:
\begin{equation}
\cos\Theta =  {(\rg-\ro) \sqrt{(\rg-m)^2\ro(\ro+2\rg)+\rg^2(\rg-3m)^2 - 4 a^2m\rg}
\over \ro^2(\rg-m) +\rp( \rp^2 -3 \rp m+2a^2)}.
\end{equation} 
Equivalently
\begin{equation}
\cos\Theta =  {(\rg-\ro) \sqrt{(\rg-m)^2(\rg-\ro)^2+4\rg(\ro-m)(\rg-m)^2 + 4\rg m (m^2-a^2)}
\over 2\rp( \rp^2 -2 \rp m+a^2) + (\ro^2-\rp^2)(\rp-m)};
\end{equation} 
With these conventions one has \, $\sign(\cos\Theta)=\sign(\rg-\ro)$. 

Note that for emission from any one of the spherical photon orbits ($r_*=r_\gamma$) we have the exact identities
\begin{equation}
\left.\sin\Theta\right|_{r_*=r_\gamma}=1; \qquad \left.\cos\Theta\right|_{r_*=r_\gamma}=0.
\end{equation}

The escape-cone solid angle is now given by
\begin{equation}
\begin{split}
\Delta\Omega_* 
& = \int_{escape~cone} \sin\vartheta \; \d\vartheta \; \d\varphi
   = \int_{-\pi}^{+\pi} \left\{1-\cos[\Theta (\Phi) ]\right\} \;\d\Phi\\
& = 2\pi - \int_{-\pi}^{+\pi} \cos[\Theta (\Phi) ]\;\d\Phi.
\end{split}
\end{equation}
By complementarity the capture-cone solid angle is
\begin{equation}
\begin{split}
\Delta\Omega_{capture}
& = \int_{capture~cone} \sin\vartheta \;\d\vartheta \; \d\varphi
   = \int_{-\pi}^{+\pi} \left\{1+\cos[\Theta (\Phi) ]\right\} \; \d\Phi\\
& = 2\pi + \int_{-\pi}^{+\pi} \cos[\Theta (\Phi) ]\;\d\Phi.
\end{split}
\end{equation}

Then by time reversal invariance the silhouette solid angle equals the capture-cone solid angle, and is given as
\begin{equation}
\begin{split}
\Delta\Omega_{silhouette} 
& = \int_{capture~cone} \sin\vartheta\; \d\varphi \; \d\vartheta 
    = \int_{-\pi}^{+\pi} \left\{1+\cos[\Theta( \Phi )]\right\} \;\d\Phi\\
& = 2\pi + \int_{-\pi}^{+\pi} \cos[\Theta( \Phi) ]\;\d\Phi.
\end{split}
\end{equation}

\enlargethispage{40pt}
As a function of $(r_*,\theta_*;\Phi)$ the locations of the spherical photon orbits, $r_\gamma$, are determined by rearranging equation (\ref{sinPhi_general}) to yield:
\begin{equation} \label{rg_general}
\rg^2(\rg-3m)+ a^2[\rg  + m + (\rg-m)\sin^2\theta_* ] = -2a\rg \sin\Phi \sin\theta_* \sqrt{\rg^2-2m\rg+a^2}.
\end{equation}
This implicitly determines $\rg(m,a;\theta_*; \Phi)$, which then in turn implicitly defines \break $\Theta(m,a;\ro,\theta_*; \Phi)$. By squaring both sides of equation (\ref{rg_general}) and subtracting we see that this is  ``just'' a sextic polynomial in $\rg$. So there is no real difficulty when throwing it at a computer for numerical analysis. However, in general this sextic has no explicit analytical solution. Due to this, one is in general unable to analytically calculate either the shape of the escape cone or the solid angle subtended thereby, instead one must appeal to numerical methods. 

The technical problem arises from the square root found on the RHS of equation \eqref{rg_general}. However, if we Taylor expand this in the slow rotation limit (i.e. $a\ll m$), this sextic equation reduces to a lower order (perturbative) equation which can then be solved explicitly; both at first and second order in the spin parameter $a$.

\section{Low-rotation limit: First-order calculation}

Let us first perform a linearized first-order calculation in the spin parameter $a$.

\subsection{Circular photon orbits}

Starting from the exact equation \eqref{rg_general}, we 
can rearrange this (still exact) as
\begin{equation}
\rg = 3m  -{2a\rg \sin\Phi \sin\theta_* \sqrt{\rg^2-2m\rg+a^2}-a^2[\rg  + m + (\rg-m)\sin^2\theta_*] \over\rg^2}.
\end{equation}
Thence, to zeroth-order in $a$ we have $\rg=3m+\O(a)$, which we then iterate to yield the first-order result:
\begin{equation}
\rg = 3m  -{2a (3m)\sin\Phi \sin\theta_* \sqrt{9m^2-6m^2} \over 9 m^2} +\O(a^2).
\end{equation}
That is
\begin{equation}
\rg = 3m  -{2\over \sqrt{3}} \; a\sin\Phi \sin\theta_*+\O(a^2).
\end{equation}
Explicitly, in terms of the dimensionless parameter $a/m$, 
\begin{equation}
\rg = 3m \left[ 1  -{2\over \sqrt{27}} \; {a\over m} \sin\Phi \sin\theta_*+\O\left(\frac{a^2}{m^2}\right) \right].
\end{equation}
Note that to first order in $a$ we see that the curve $\rg (\Phi) $ is a \emph{lima\c{c}on}. 

\subsection{Shape of the escape cone}

We can now [from equation \eqref{sintheta_general}] explicitly calculate the escape cone opening angle (to first order in $a$):
\begin{equation} \label{sintheta_1st_order}
\sin\Theta  = 3\sqrt{3} \, {m\over \ro} \sqrt{1-\frac{2m}{\ro}} - 3\left(1-\frac{9m^2}{\ro^2}\right)  \sin\Phi \sin\theta_* \sqrt{1-\frac{2m}{\ro}} \;{a\over\ro}
+ \O(a^2) .
\end{equation}
To first order in $a$ the curve $[\sin\Theta](\Phi) $ is again a \emph{lima\c{c}on}. 

\enlargethispage{20pt}
Thence
{\small
\begin{equation}\label{costheta_1st_order}
\cos\Theta = -\left(1-\frac{3m}{\ro}\right) \sqrt{1+\frac{6m}{\ro}} 
- 9 \sqrt{3} \sin\Phi\sin\theta_* {(1+3m/\ro)(1-2m/\ro) \over \sqrt{1+6m/\ro}} {ma\over \ro^2} 
+\O(a^2).
\end{equation}
}\!\!
To first order in $a$ the curve $[\cos\Theta](\Phi) $ is again a \emph{lima\c{c}on}. 

In summary\enlargethispage{30pt}
\begin{equation}
\sin\Theta =  3\sqrt{3} \, {m\over \ro} \sqrt{1-\frac{2m}{\ro}}\left[
 1 - {1\over\sqrt{3}}\left(1-\frac{9m^2}{\ro^2}\right)  \sin\Phi \sin\theta_* \;{a\over m}
+ \O\left(\frac{a^2}{m^2}\right)
\right];
\end{equation}
and
\begin{eqnarray}
\cos\Theta&=&  -\left(1-\frac{3m}{\ro}\right) \sqrt{1+\frac{6m}{\ro}} 
\nonumber\\
&&\times 
\left[ 1 +  9 \sqrt{3} \sin\Phi\sin\theta_*\;  {1+3m/\ro\over 1-3m/\ro} {1-2m/\ro\over 1+6m/\ro}\; {ma\over \ro^2} 
+\O\left(a^2\over m^2\right)
\right].
\end{eqnarray}

Then in terms of the zeroth-order Schwarzschild results
\begin{equation}
\sin\Theta = \sin\Theta_0 \left[
 1 - {1\over\sqrt{3}}\left(1-\frac{9m^2}{\ro^2}\right)  \sin\Phi \sin\theta_* \;{a\over m}
+ \O\left(\frac{a^2}{m^2}\right)
\right];
\end{equation}
and
\begin{equation}
\cos\Theta = \cos\Theta_0 
\left[ 1 +  9 \sqrt{3} \sin\Phi\sin\theta_*\;  {1+3m/\ro\over 1-3m/\ro} {1-2m/\ro\over 1+6m/\ro}\; {ma\over \ro^2} 
+\O\left(a^2\over m^2\right)
\right].
\end{equation}
Note that at first order in $a$ this distorts the \emph{shape} of the escape cone away from the Schwarzschild result. 

\subsection{Solid angle subtended by the escape cone}

However for the solid angle subtended by the escape cone we have
\begin{equation}
\begin{split}
\Delta\Omega_* = \bigoint & \left[ 1  + \left(1-\frac{3m}{\ro}\right) \sqrt{1+\frac{6m}{\ro}}\right. \\
& \hspace{4.5mm} \left. + 9 \sqrt{3} \sin\Phi\sin\theta_* {(1+3m/\ro)(1-2m/\ro) \over \sqrt{1+6m/\ro}} {ma\over \ro^2} 
+\O(a^2)\right] d\Phi.
\end{split}
\end{equation}
That is
\begin{equation}\label{Omega_1st_order}
\Delta\Omega_* = 2\pi\left[ 1 + \left(1-\frac{3m}{\ro}\right) \sqrt{1+\frac{6m}{\ro}} 
+\O(a^2)\right] =  (\Delta\Omega_*)_0 + \O(a^2).
\end{equation}
We see that the $\O(a)$ contribution to the escape cone solid angle vanishes after integration. (With hindsight this is obvious --- by symmetry $\Delta\Omega_*$ cannot depend on the sense of rotation, and must be an even function of $a$.)  

By symmetry, for the capture cone we have 
\begin{equation}
\Delta\Omega_{capture} = 2\pi\left[ 1 - \left(1-\frac{3m}{\ro}\right) \sqrt{1+\frac{6m}{\ro}} 
+\O(a^2)\right].
\end{equation}
By running the captured null geodesics backwards in time this tells us that the solid angle subtended by the black hole silhouette is also 
\begin{equation}
\Delta\Omega_{silhouette} = 2\pi\left[ 1 - \left(1-\frac{3m}{\ro}\right) \sqrt{1+\frac{6m}{\ro}} 
+\O(a^2)\right].
\end{equation}

Overall, at first-order, while the \emph{shape} of the escape cone and silhouette are certainly distorted, the solid angle receives no corrections at this order. Hence, we are required to conduct a second-order calculation to see nontrivial changes to the solid angle which we shall do after looking at a few special cases. 

\subsection{Some special cases}
As a consistency check it is a good idea to consider three special cases:
\begin{enumerate}
\item Near horizon emission $r_* - 2m \ll m$.
\item Emission from near the naive photon sphere $r_*\approx 3m$.
\item Long distance reception $r_* \gg m$. 
\end{enumerate}
The first situation corresponds to photons escaping from just above the horizon, which is perturbatively located  at
\begin{equation} 
r_H = m+\sqrt{m^2-a^2} = 2m \left[1 + \O\left(\frac{a^2}{m^2}\right) \right].
\end{equation}
So to the required order of accuracy we can simply set $r_H\to 2m$.

In this first situation we expect the escape cone to be extremely small and narrow.
The second case ($r_* \approx 3m$) corresponds to a close to 50-50 split between escape and capture. 
These first two cases are particularly important for understanding energy fluxes and energy balance in the near-horizon environment~\cite{thermalizing, thermal-SgA}.

The third case ($r_* \gg m$) is somewhat different in character, being relevant to the practical determination of the astronomically important Kerr silhouette. 
In this situation we expect the capture cone to be extremely small and narrow.

\subsubsection{Near horizon behaviour}
By Taylor expanding equations \eqref{sintheta_1st_order}, \eqref{costheta_1st_order}, and \eqref{Omega_1st_order} around $r=2m$, we get
\begin{equation}
\sin\Theta = 
{\sqrt{27}\over 2} \sqrt{r_*-2m\over 2m} \left[ 1 + {5 \sin\Phi\sin\theta_*\over 4 \sqrt{3}} {a\over m} 
+ \O\left(\frac{a^2}{m^2}\right) + \O\left(r_*-2m\over 2m\right) \right].
\end{equation}
Thence 
\begin{equation}
\cos\Theta = 1 -\left({27\over8} + {45\sqrt{3} \sin\theta_*\sin\Phi\over 16} {a\over m} \right){r-2m\over 2m}
+ \O\left(\frac{a^2}{m^2}\right) + \O\left([r_*-2m]^2\over m^2\right).
\end{equation}

Then in this regime, after integration over $\Phi$, we see that for the escape cone solid angle
\begin{equation}
\Delta\Omega_* =  {27\pi\over 8}\;{r_*-2m\over m} + \O\left(\frac{a^2}{m^2}\right) + \O\left([r_*-2m]^2\over m^2\right).
\end{equation}
We note that the escape cone is indeed asymptotically narrow as the emission point approaches the horizon, and there is no $\O(a)$ contribution.

\subsubsection{Naive photon sphere}
By Taylor expanding equations \eqref{sintheta_1st_order}, \eqref{costheta_1st_order} and \eqref{Omega_1st_order} around the zeroth-order photon sphere at $r=3m$, we get
\begin{equation}
\sin\Theta = 1 - \frac{2\sin \Phi \sin \theta_*}{3\sqrt{3}}\frac{a}{m^2}(r_*-3m) + \O\left(\frac{a^2}{m^2}\right) + \O\left([r_*-3m]^2\over [3m]^2\right).
\end{equation}
Thence
\begin{equation}
\begin{split}
\cos\Theta & = \frac{2\sin \Phi \sin \theta_*}{3}\frac{a}{m}+\left(\frac{1}{\sqrt{3}m}-\frac{\sin \Phi \sin \theta_*}{27}\frac{a}{m^2}\right)(r_*-3m) \\
& \hspace{4.5mm} + \O\left(\frac{a^2}{m^2}\right) + \O\left([r_*-3m]^2\over [3m]^2\right),
\end{split}
\end{equation}
and in this regime
\begin{equation}
\Delta\Omega_* = 2\pi \left[ 1 + \frac{1}{\sqrt{3}m}(r_*-3m) + \O\left(\frac{a^2}{m^2}\right) + \O\left([r_*-3m]^2\over [3m]^2\right) \right],
\end{equation}
implying 
\begin{equation}
\Delta\Omega_{silhouette} = 2\pi \left[ 1 - \frac{1}{\sqrt{3}m}(r_*-3m) + \O\left(\frac{a^2}{m^2}\right) + 
\O\left([r_*-3m]^2\over [3m]^2\right) \right].
\end{equation}
Again, we see that the solid angle receives no first order correction. 

\subsubsection{Large distance behaviour}
By Taylor expanding equations \eqref{sintheta_1st_order}, \eqref{costheta_1st_order}, and \eqref{Omega_1st_order} around spatial infinity $r_*=\infty$, we get
\begin{equation}
\sin\Theta = 
{\sqrt{27}} \,{m\over r_*} \left[ 1 - {\sin\Phi\sin\theta_*\over  \sqrt{3}} {a\over m} 
+ \O\left(\frac{a^2}{m^2}\right) + \O\left(m\over r_*\right) \right].
\end{equation}
Thence \enlargethispage{40pt}
\begin{equation}
\cos\Theta = -1 + {27 m^2\over 2 r_*^2} - 9\sqrt{3}\sin\theta_*\sin\Phi\;{am\over r_*^2}
+\O\left(\frac{a^2}{m^2}\right) + \O\left(m^3\over r_*^3\right).
\end{equation}

Then in this regime, after integration over $\Phi$, we see that for the silhouette solid angle
\begin{equation}
\Delta\Omega_{silhouette} = {27 \pi m^2\over  r_*^2} +\O\left(\frac{a^2}{m^2}\right) + \O\left(m^3\over r_*^3\right).
\end{equation}
As expected, the capture cone is in this situation asymptotically small.

Note that all of these three results are compatible with those obtained for the Schwarzschild spacetime in equation (\ref{E:Sch-special}).

\section{Low-rotation limit: Second-order calculation}

The second-order calculation is slightly tedious, and somewhat subtle, but is not intrinsically difficult.

\subsection{Circular photon orbits}

We start by making an ansatz for $\rg$
\begin{equation}
\rg = 3m \left\{ 1  +Q_1  {a\over m}
+ Q_2 {a^2\over m^2} + \O\left(\frac{a^3}{m^3}\right) \right\}.
\end{equation}
Now insert this ansatz into equation \eqref{rg_general}, 
and Taylor expand to 3rd order in $a$.  Collecting the terms in $a$ and $a^2$ yields
\begin{equation}
Q_1 = -{2\over\sqrt{27} }\sin\Phi \sin\theta_* ,
\end{equation}
and 
\begin{equation}
Q_2 = -{2\over27}(2+\sin^2\theta_* [1-2\sin^2\Phi]) =  -{2\over27}(2+\sin^2\theta_* ) +Q_1^2.
\end{equation}
Note that
\begin{equation}
\oint \sin\Phi \; \d\Phi=0; \qquad {1\over2\pi} \oint \sin^2\Phi \; \d\Phi={1\over2};
\end{equation}
which implies
\begin{equation}
{1\over 2\pi} \oint Q_1 \; \d\Phi=0; \qquad {1\over2\pi} \oint Q_1^2 \; \d\Phi= {2\over27} \sin^2\theta_* ; 
\qquad {1\over2\pi} \oint Q_2  \; \d\Phi=-{4\over27}.
\end{equation}
That is
\begin{equation}
\langle \rg \rangle := {1\over 2\pi} \oint r_\gamma \; \d\Phi =  3m \left\{ 1-{4\over27} {a^2\over m^2} + \O\left(\frac{a^3}{m^3}\right) \right\}.
\end{equation}
So ``on average'', at least for slow rotation, the spherical photon orbits do not move too far from the Schwarzschild result $r_\gamma=3m$. Similarly
\begin{equation}
\langle \rg^2 \rangle := {1\over 2\pi} \oint r_\gamma^2 \; \d\Phi = 
 (3m)^2 \left\{ 1 +[\langle Q_1^2\rangle + 2 \langle Q_2\rangle]  {a^2\over m^2} + \O\left(\frac{a^3}{m^3}\right) \right\},
\end{equation}
implying
\begin{equation}
\langle \rg^2 \rangle = 
 (3m)^2 \left\{ 1 +\left[{2\over27} \sin^2\theta_* -{8\over27} \right]  {a^2\over m^2} + \O\left(\frac{a^3}{m^3}\right) \right\}.
\end{equation}
Thence we see
\begin{equation}
\langle \rg^2 \rangle - \langle \rg \rangle^2 = 
 m^2 \left\{\left[{2\over3} \sin^2\theta_*  \right]  {a^2\over m^2} + \O\left(\frac{a^3}{m^3}\right) \right\}.
\end{equation}
So the ``standard deviation'' in the radius $\rg$ of the spherical  photon orbits is
\begin{equation}
\sqrt{\langle \rg^2 \rangle - \langle \rg \rangle^2} = 
 a \sqrt{2\over3} \; \sin\theta_* \left\{1 + \O\left(\frac{a}{m}\right) \right\}.
\end{equation}
In fact, compared to the zeroth-order photon sphere at $3m$ one has
\begin{equation}
{\sqrt{\langle \rg^2 \rangle - \langle \rg \rangle^2} \over 3m}= 
 {a\over m} \sqrt{2\over27} \; \sin\theta_* \left\{1 + \O\left(\frac{a}{m}\right) \right\} 
 < {a\over m} \left\{1 + \O\left(\frac{a}{m}\right) \right\} .
\end{equation}

\subsection{Shape of the escape cone}
\subsubsection{Evaluating $\sin(\Theta)$}

To find the shape of the second order escape cone we find it most useful to separate the explicit $a$ dependence in $\sin\Theta$, which only shows up at second order, from the implicit $a$ dependence arising from $\rg(a)$, which has contributions arising at both first and second order:
\begin{equation}
\rg(a) = 3m \left\{ 1  +Q_1  {a\over m}
+ Q_2 {a^2\over m^2} + \O\left(\frac{a^3}{m^3}\right) \right\}.
\end{equation}
Let us first, using the zeroth-order  approximation $\rg\approx3m$ for the photon orbits, define
\begin{equation}
\sin\Theta_{analytic}= \left.\sin\Theta\right|_{\rg=3m} = {3\sqrt{3m^2+a^2} \sqrt{r_*^2-2mr_*+a^2} \over r_*^2+3a^2},
\end{equation}
and
\begin{equation}
\cos\Theta_{analytic}= \left.\cos\Theta\right|_{\rg=3m} =-  {(r_*-3m) \sqrt{r_*^2+6mr_*-3a^2} \over r_*^2+3a^2}.
\end{equation}
These quantities are approximations that clearly have the appropriate Schwarzschild limit as $a\to0$, but have the significant technical advantage that they carry some non-perturbative $a$ dependence, coming from the explicit $a$ dependence in the expressions for $\sin\Theta$ and $\cos\Theta$. 

In particular $\sin\Theta_{analytic}$ vanishes on the exact horizon $r_H = m+\sqrt{m^2-a^2}$, as it should, and also vanishes for asymptotically large $r_*$, as it should. 
Furthermore $\cos\Theta_{analytic}$ vanishes on the naive photon sphere $r_*\to3m$. So these two quantities, $\sin\Theta_{analytic}$ and $\cos\Theta_{analytic}$, capture key features of the exact analytic results and can be used as the basis for developing useful approximations.
Finally we note that the two ratios $\sin\Theta/\sin\Theta_{analytic}$ and $\cos\Theta/\cos\Theta_{analytic}$ prove relatively tractable to work with.

Specifically, inserting $\rg(a)$ into equation \eqref{sintheta_general} for $\sin(\Theta)$ and Taylor expanding $\sin\Theta/\sin\Theta_{analytic}$ to second order in $a$ (including both the explicit and the implicit dependence on $a$) yields:
\begin{equation}\label{sintheta_2nd_order}
\sin\Theta =  \sin\Theta_{analytic}
\left\{ 1+  k_1 Q_1  {a\over m}+ [ k_1 Q_2 +k_2 Q_1^2 ] {a^2\over m^2} + \O\left(\frac{a^3}{m^3}\right) \right\}.
\end{equation} 
Here the dimensionless coefficients $k_1$ and $k_2$ are given by  simple polynomials:
\begin{equation}
k_1 = {3\over2} \left[1-\left(\frac{3m}{r_*}\right)^2\right];
\qquad
k_2 = 
- {3\over 4} \left[1+\left(\frac{6m}{r_*}\right)^2-3\left(\frac{3m}{r_*}\right)^4\right].
\end{equation}

Note all the angular dependence is hidden in $Q_1$ and $Q_2$,
\begin{equation}
Q_1 = -{2\over\sqrt{27} }\sin\Phi \sin\theta_* ,
\qquad
Q_2 = -{2\over27}(2+\sin^2\theta_* [1-2\sin^2\Phi]).
\end{equation}
The two dimensionless coefficients  $k_i$ have no angular dependence. 
This expression then explicitly gives the \emph{shape} of the escape cone, for emission from the point $(r_*,\theta_*)$, as a function of the free parameter $\Phi$ present in $Q_1$ and $Q_2$.
Indeed, concentrating on the $\Phi$ dependence we can write
\begin{equation}
[\sin\Theta](\Phi) = S_0 + S_1 \sin\Phi + S_2 \sin^2\Phi + ...
\end{equation}
This curve no longer qualifies as a \emph{lima\c{c}on}, though it might resonably be considered a generalization thereof. 
 
\subsubsection{Evaluating $\cos(\Theta)$}
 
 Now consider $\cos\Theta$, Taylor expanding $\cos\Theta/\cos\Theta_{analytic}$ we find
 \begin{equation}\label{costheta_2nd_order}
\cos\Theta =  \cos\Theta_{analytic}
\left\{ 1+  \tilde k_1 Q_1  {a\over m}+ [ \tilde k_1 Q_2 +\tilde k_2 Q_1^2 ] {a^2\over m^2} + \O\left(\frac{a^3}{m^3}\right) \right\}.
\end{equation} 
Here the two dimensionless coefficients $\tilde k_1$ and $\tilde k_2$ are now given by somewhat messier expressions
\begin{equation}
\tilde k_1 = -{81\over2} {(1-2m/r_*)(1+3m/r_*)\over(1-3m/r_*)(1+6m/r_*)} \; {m^2\over r_*^2} ;
\end{equation}
\begin{equation}
\tilde k_2 = 
- {81\over 8} {(1-2m/r_*)
\left[1+3\left(3m\over r_*\right)-2\left(3m\over r_*\right)^2-18\left(3m\over r_*\right)^3-12\left(3m\over r_*\right)^4\right]
\over(1-3m/r_*)(1+6m/r_*)^2}{m^2\over r_*^2}.
\end{equation}
Note all the angular dependence is again hidden in $Q_1$ and $Q_2$,
and that again the two dimensionless coefficients  $\tilde k_i$ have no angular dependence. 
While the $\tilde k_i$ naively possess poles at $r_*=3m$ these poles are cancelled by the explicit zero in $\cos\Theta_{analytic}$, so it is worthwhile to rewrite $\cos\Theta$ in terms of the manifestly finite polynomial expressions
\begin{equation}
\bar k_1 = -{81\over2} \; \left[1+{3m\over r_*}\right] \left[1+{6m\over r_*}\right];
\end{equation}
\begin{equation}
\bar k_2 = 
- {81\over 8} 
\left[1+3\left(3m\over r_*\right)-2\left(3m\over r_*\right)^2-18\left(3m\over r_*\right)^3-12\left(3m\over r_*\right)^4\right];
\end{equation}
and
 \begin{eqnarray}
 \label{costheta_2nd_order_B}
\cos\Theta &=&  \cos\Theta_{analytic} -
{\sqrt{1+6m/r_*-3a^2/r_*^2} \over 1+3a^2/r_*^2}  {(1-2m/r_*)\over(1+6m/r_*)^2}{m^2\over r_*^2} 
\nonumber\\
&&\qquad\qquad
\times
\left\{\bar k_1 Q_1  {a\over m}+ [ \bar k_1 Q_2 +\bar k_2 Q_1^2 ] {a^2\over m^2} + \O\left(\frac{a^3}{m^3}\right) \right\}.
\end{eqnarray} 
Indeed, concentrating on the $\Phi$ dependence hiding in $Q_1$ and $Q_2$ we can write
\begin{equation}
[\cos\Theta](\Phi) = C_0 + C_1 \sin\Phi + C_2 \sin^2\Phi + ...
\end{equation}
This curve no longer qualifies as a \emph{lima\c{c}on}, though it might reasonably be considered a generalization thereof. 
 
\subsection{Solid angle subtended by the escape cone}

To calculate the escape cone solid angle we integrate over $\Phi$, noting that
\begin{equation}
\oint Q_1 \; \d\Phi=0; \qquad \oint Q_2 \; \d\Phi=-{8\pi\over27}; \qquad \hbox{and} \qquad \oint Q_1^2 \; \d\Phi= {4\pi\over27}\sin^2\theta_*.
\end{equation}
Then we have
\begin{eqnarray}
\oint \cos\Theta \; \d\Phi & =& 2\pi \cos\Theta_{analytic}
-  {\sqrt{1+6m/r_*-3a^2/r_*^2} \over 1+3a^2/r_*^2}{(1-2m/r_*)\over(1+6m/r_*)^2}{m^2\over r_*^2} 
\nonumber\\
&&
\qquad\qquad \times
\left\{
\left[ -{8\pi\over27} \bar k_1 + {4\pi\over27}\sin^2\theta_*\bar k_2 \right] {a^2\over m^2} 
+ \O\left(\frac{a^3}{m^3}\right) \right\}.
\end{eqnarray}

At this stage it becomes useful to change the normalizations and define
\begin{equation}
\hat k_1 =\left[1+{3m\over r_*}\right] \left[1+{6m\over r_*}\right];
\end{equation}
\begin{equation}
\hat k_2 = 
\left[1+3\left(3m\over r_*\right)-2\left(3m\over r_*\right)^2-18\left(3m\over r_*\right)^3-12\left(3m\over r_*\right)^4\right];
\end{equation}
so that 
\begin{eqnarray}
\oint \cos\Theta \; \d\Phi & =& 2\pi \cos\Theta_{analytic}
+  {3\pi\over2} {\sqrt{1+6m/r_*-3a^2/r_*^2} \over 1+3a^2/r_*^2}{(1-2m/r_*)\over(1+6m/r_*)^2}{m^2\over r_*^2} 
\nonumber\\
&&
\qquad\qquad \times
\left\{
\left[ -8 \hat k_1 + \sin^2\theta_*\hat k_2 \right] {a^2\over m^2} 
+ \O\left(\frac{a^3}{m^3}\right) \right\}.
\end{eqnarray}

Thence for the escape cone solid angle
\begin{equation}
\Delta\Omega_* = \oint [1-\cos\Theta]\d\Phi = 2\pi - \oint \cos\Theta\,\d\Phi,
\end{equation}
we finally have
\begin{eqnarray}
\Delta\Omega_* & =& 2\pi [1-\cos\Theta_{analytic}] 
-{3\pi\over2}  {\sqrt{1+6m/r_*-3a^2/r_*^2} \over 1+3a^2/r_*^2}{(1-2m/r_*)\over(1+6m/r_*)^2}{m^2\over r_*^2} 
\nonumber\\
&& \qquad\qquad \times
\left\{ 
\left[ -{8} \hat k_1 +\sin^2\theta_*\hat k_2 \right] {a^2\over m^2} 
+ \O\left(\frac{a^3}{m^3}\right) \right\}.
\end{eqnarray}

\enlargethispage{30pt}
In view of the previous argument that $\Delta\Omega_* $ should be an even function of $a$ this can actually be slightly strengthened
\begin{eqnarray}
\label{Omega_2nd_order_B}
\Delta\Omega_* & =& 2\pi \left[1+{(r_*-3m)\sqrt{r_*^2+6mr_*-3a^2} \over r_*^2+3a^2}\right] 
\nonumber\\
&&
\qquad
 - {3\pi\over2} {\sqrt{1+6m/r_*-3a^2/r_*^2} \over 1+3a^2/r_*^2}{(1-2m/r_*)\over(1+6m/r_*)^2}{m^2\over r_*^2} 
 \nonumber\\
 &&\qquad\qquad\times
\left\{ 
\left[ -8 \hat k_1 + \sin^2\theta_*\hat k_2 \right] {a^2\over m^2} 
+ \O\left(\frac{a^4}{m^4}\right) \right\}.
\end{eqnarray}
Note this is manifestly a function of $a^2$ and has all the appropriate limits.

\clearpage
\subsection{Some special cases}

As a consistency check it is a good idea to look at three special cases:
\begin{enumerate}
\item Near horizon emission $r_* - r_H \ll m$.
\item Emission from near the naive photon sphere $r_*\approx 3m$.
\item Long distance reception $r_* \gg m$. 
\end{enumerate}
The first situation corresponds to photons escaping from just above the horizon, now perturbatively located  at
\begin{equation} 
r_H = m+\sqrt{m^2-a^2} = 2m \left[1-{1\over4} {a^2\over m^2} + \O\left(\frac{a^4}{m^4}\right) \right].
\end{equation}
In view of this it is no longer appropriate to simply set $r_H \to 2m$, more care must be taken.
The third case is relevant to determining the Kerr silhouette. 
In these three special cases the functions $\sin\Theta_0$, $\cos\Theta_0$, and the three functions $k_i(r_*)$, all  simplify.

\subsubsection{Near horizon behaviour}
Observe that near the horizon
\begin{equation}
k_1(r_*) = {3\over2} \left[1-\left(\frac{3m}{r_H}\right)^2\right] + \O\left(r_*-r_H\over r_H\right);
\qquad
\end{equation}
\begin{equation}
k_2(r_*) = 
- {3\over 4} \left[1+\left(\frac{6m}{r_H}\right)^2-3\left(\frac{3m}{r_H}\right)^4\right] + \O\left(r_*-r_H\over r_H\right).
\end{equation}
So to the required order of accuracy in the near-horizon regime we have
\begin{eqnarray}\label{sintheta_2nd_order-near-horizon}
\sin\Theta =  \sin\Theta_{analytic} &\times& 
\left\{ 1+  k_1(r_H) Q_1  {a\over m}+ [ k_1(r_H) Q_2 +k_2(r_H) Q_1^2 ] {a^2\over m^2} 
\right.
\nonumber\\
&& \qquad\qquad
\left.+ \O\left(\frac{a^3}{m^3}\right) 
+ \O\left(r_*-r_H\over r_H\right) \right\}.
\end{eqnarray} 

\enlargethispage{20pt}
To be a little more explicit
\begin{eqnarray}\label{sintheta_2nd_order-near-horizon2}
\sin\Theta &=&  {3\sqrt{3m^2+a^2} \sqrt{r_*^2-2mr_*+a^2} \over r_*^2+3a^2}
\nonumber\\
&& \times
\left\{ 1+  k_1(r_H) Q_1  {a\over m}+ [ k_1(r_H) Q_2 +k_2(r_H) Q_1^2 ] {a^2\over m^2} 
\right.
\nonumber\\
&& 
\qquad\qquad\qquad
\left.+ \O\left(\frac{a^3}{m^3}\right) 
+ \O\left(r_*-r_H\over r_H\right) \right\}.
\end{eqnarray} 
On the horizon itself $\sin\Theta\to0$, and sufficiently near the horizon $\sin\Theta=\O\left(\sqrt{r_*-r_H}\right)$. 

For the quantity $\cos\Theta$ we note that
\begin{equation}
\bar k_1(r_*) = -{81\over2} \; \left[1+{3m\over r_H}\right] +\O\left(r_*-r_H\over r_H\right); 
\end{equation}
\begin{equation}
\bar k_2(r_*) = 
- {81\over 8} 
\left[1+3\left(3m\over r_H\right)-2\left(3m\over r_H\right)^2-18\left(3m\over r_H\right)^3-12\left(3m\over r_H\right)^4\right] 
+\O\left(r_*-r_H\over r_H\right); 
\end{equation}
and find the relatively messy result
 \begin{eqnarray}
 \label{costheta_2nd_order_B2}
\cos\Theta &=&  \cos\Theta_{analytic} - {\sqrt{1+6m/r_*-3a^2/r_*^2} \over 1+3a^2/r_*^2} 
{(1-2m/r_*)\over(1+6m/r_*)^2}{m^2\over r_*^2} 
\nonumber\\
&&\qquad\qquad
\times
\left\{\bar k_1(r_H) Q_1  {a\over m}+ [ \bar k_1(r_H) Q_2 +\bar k_2(r_H)  Q_1^2 ] {a^2\over m^2} 
\right.\nonumber\\
&& \qquad\qquad\qquad
\left.+ \O\left(\frac{a^3}{m^3}\right) +\O\left(r_*-r_H\over r_H\right) \right\}.
\end{eqnarray} 
Exactly on the horizon we know the exact result is $\cos\Theta\to-1$ but the perturbative calculation merely yields $\cos\Theta\to -1 +\O(a^2/m^2)$.

For the escape cone solid angle
\begin{eqnarray}
\label{Omega_2nd_order_B2}
\Delta\Omega_* & =& 2\pi \left[1+{(r_*-3m)\sqrt{r_*^2+6mr_*-3a^2} \over r_*^2+3a^2}\right] 
\nonumber\\
&&
\qquad
 - 2\pi {\sqrt{1+6m/r_*-3a^2/r_*^2} \over 1+3a^2/r_*^2}{(1-2m/r_*)\over(1+6m/r_*)^2}{m^2\over r_*^2} 
 \nonumber\\
 &&\qquad\qquad\times
\left\{ 
\left[ {4\over27} \bar k_1(r_H) - {2\over27}\sin^2\theta_*\bar k_2(r_H) \right] {a^2\over m^2} \right.
\nonumber\\
 &&\qquad\qquad\qquad
\left.
+ \O\left(\frac{a^4}{m^4}\right) +\O\left(r_*-r_H\over r_H\right)\right\}.
\end{eqnarray}
If we use the rescaled quantities $\hat k_1$ and $\hat k_2$ then
\begin{eqnarray}
\label{Omega_2nd_order_B22}
\Delta\Omega_* & =& 2\pi \left[1+{(r_*-3m)\sqrt{r_*^2+6mr_*-3a^2} \over r_*^2+3a^2}\right] 
\nonumber\\
&&
\qquad
 - 2\pi {\sqrt{1+6m/r_*-3a^2/r_*^2} \over 1+3a^2/r_*^2}{(1-2m/r_*)\over(1+6m/r_*)^2}{m^2\over r_*^2} 
 \nonumber\\
 &&\qquad\qquad\times
\left\{ 
\left[ -6 \hat k_1(r_H) - {3\over4}\sin^2\theta_*\hat k_2(r_H) \right] {a^2\over m^2} \right.
\nonumber\\
 &&\qquad\qquad\qquad
\left.
+ \O\left(\frac{a^4}{m^4}\right) +\O\left(r_*-r_H\over r_H\right)\right\}.
\end{eqnarray}

Exactly on the horizon we know the exact result is $\Delta\Omega_*\to0$ but the perturbative calculation merely yields $\Delta\Omega_*\to \O(a^2/m^2)$.

\subsubsection{Naive photon sphere}
On the naive photon sphere at $r_*=3m$ we have
\begin{equation}
\sin\Theta_{analytic}\to 1; \qquad k_1 \to 0; \qquad k_2 \to -{3\over 2}.
\end{equation}
Consequently
\begin{equation}
\sin\Theta \to 1 - {3\over2} Q_1^2 {a^2\over m^2} +\O\left(a^3\over m^3\right).
\end{equation}

Similarly
\begin{equation}
\cos\Theta_{analytic}\to 0; \qquad \bar k_1 \to -243; \qquad \bar k_2 \to {567\over 2} = {7\over2} \; 81;
\end{equation}
and
\begin{equation}
{\sqrt{1+6m/r_*-3a^2/r_*^2} \over 1+3a^2/r_*^2}  {(1-2m/r_*)\over(1+6m/r_*)^2}{m^2\over r_*^2} 
\to {\sqrt{3}\over243} {\sqrt{1-{a^2\over9m^2}}\over1+ {a^2\over 9 m^2}} = {1\over 81 \sqrt{3}} +\O\left(a^2\over m^2\right).
\end{equation}

Consequently
 \begin{eqnarray}
 \label{costheta_2nd_order_B22}
\cos\Theta &\to&  -\sqrt{3}
\left\{ Q_1  {a\over m}+\left[  Q_2 +{7\over6}  Q_1^2 \right] {a^2\over m^2} + \O\left(\frac{a^3}{m^3}\right) \right\}.
\end{eqnarray} 
As expected, at the naive photon sphere we see $\cos\Theta\to \O(a/m)$.

\enlargethispage{20pt}
For the escape cone solid angle, integrating over $\Phi$,  we have
\begin{eqnarray}
\label{Omega_2nd_order_BBB}
\Delta\Omega_* & =& 2\pi 
 - 2\pi {\sqrt{3}}
\left\{ 
\left[ {4\over27} - {7\over81}\sin^2\theta_* \right] {a^2\over m^2} 
+ \O\left(\frac{a^4}{m^4}\right) \right\},
\end{eqnarray}
Note this is $\Delta\Omega_* = 2\pi +\O(a^2/m^2)$, close to a 50-50 split, as expected. 

\subsubsection{Large distance behaviour}
Finally at large distances we note
\begin{equation}
\sin\Theta_{analytic}\to {3\sqrt{3m^2+a^2}\over r_* } +\O(m^2/r_*^2);
\end{equation}
while
\begin{equation}
k_1 = {3\over2}+\O(m^2/r_*^2); \qquad k_2 = -{3\over 4}+\O(m^2/r_*^2)
\end{equation}
Thence
\begin{equation}
\sin\Theta = {3\sqrt{3m^2+a^2}\over r }\left[ 1 +{3\over2} Q_1 {a\over m} +\left[{3\over2}Q_2-{3\over4} Q_1^2 \right]{a^2\over m^2} + \O(a^3/m^3)+\O(m/r) \right]
\end{equation}

Similarly
\begin{equation}
\cos\Theta_{analytic}\to 1 - {9\over2}\;{3m^2+a^2\over r_*^2} +\O(m^3/r_*^3);
\end{equation}
while
\begin{equation}
\bar k_1 = -{81\over2}+\O(m/r); \qquad k_2 = -{81\over 8}+\O(m/r);
\end{equation}
and
\begin{equation}
{\sqrt{1+6m/r_*-3a^2/r_*^2} \over 1+3a^2/r_*^2}  {(1-2m/r_*)\over(1+6m/r_*)^2}{m^2\over r_*^2} 
\to {m^2\over r^2}  \left[1+\O\left(m\over r\right) \right].
\end{equation}

Thence
 \begin{eqnarray}
 \label{costheta_2nd_order_B7}
\cos\Theta &=&  1 - {9\over2}\;{3m^2+a^2\over r_*^2} + {81\over2} {m^2\over r_*^2} 
\left\{ Q_1  {a\over m}+ \left[ Q_2 +{1\over 4}  Q_1^2 \right] {a^2\over m^2} 
\right. \nonumber\\
&&\qquad\qquad\qquad\qquad\qquad\qquad \left.
+ \O\left(\frac{a^3}{m^3}\right) +\O\left(m\over r_*\right)\right\}.
\end{eqnarray} 

To obtain the escape cone solid  angle we again integrate over $\Phi$ obtaining
 \begin{eqnarray}
 \label{costheta_2nd_order_B7b}
 \Delta\Omega_* 
&=&  {9\pi }\;{3m^2+a^2\over r_*^2} - {81\over2} {m^2\over r_*^2} 
\left\{ \oint \left[ Q_2 +{1\over 4}  Q_1^2 \right] \d \Phi {a^2\over m^2} 
+ \O\left(\frac{a^3}{m^3}\right) +\O\left(m\over r_*\right)\right\}.
\nonumber\\
&&
\end{eqnarray} 

Thence
\begin{eqnarray}
 \label{costheta_2nd_order_B7c}
 \Delta\Omega_* 
&=&  {9\pi }\;{3m^2+a^2\over r_*^2} - {3\pi\over2}  {m^2\over r_*^2} 
\left\{ \left[ -8 +\sin^2\theta_* \right]  {a^2\over m^2} 
+ \O\left(\frac{a^3}{m^3}\right) +\O\left(m\over r_*\right)\right\}.
\nonumber\\
&&
\end{eqnarray} 

That is
\begin{eqnarray}
 \label{costheta_2nd_order_B7d}
 \Delta\Omega_* 
&=&  {27\pi m^2\over r_*^2} +{3\pi\over2}  {m^2\over r_*^2} 
\left\{ \left[ +14 -\sin^2\theta_* \right]  {a^2\over m^2} 
+ \O\left(\frac{a^3}{m^3}\right) +\O\left(m\over r_*\right)\right\}.
\nonumber\\
&&
\end{eqnarray} 

Note that all of these three results are compatible with those obtained for the Schwarzschild spacetime in equation (\ref{E:Sch-special}).

\section{Conclusions}
While the escape cones for Kerr in general are intractable in an analytic sense, in the slow rotation limit, closed form expressions for the escape cones come quite readily. In first order in $a$ we find that the escape cone deviates away from the Schwarzschild result and is given by the \emph{lima\c{c}on}:

\begin{equation}
\sin\Theta = 3\sqrt{3} \, {m\over \ro} \sqrt{1-\frac{2m}{\ro}}\left[
 1 - {1\over\sqrt{3}}\left(1-\frac{9m^2}{\ro^2}\right)  \sin\Phi \sin\theta_* \;{a\over m}
+ \O\left(\frac{a^2}{m^2}\right)
\right].
\end{equation}

However, we see that the solid angle is given by
\begin{equation}
\Delta\Omega_* = 2\pi\left[ 1 + \left(1-\frac{3m}{\ro}\right) \sqrt{1+\frac{6m}{\ro}} 
+\O(a^2)\right],
\end{equation}
the order $a$ contribution vanishes after integration and hence the solid angle receives no contributions at this order.

\enlargethispage{20pt}
To see any contributions to the solid angle, we need to conduct a second order analysis. Here we find
\begin{equation}
\sin\Theta =  \sin\Theta_{analytic} 
\left\{ 1+  k_1 Q_1  {a\over m}+ [ k_1 Q_2 +k_2 Q_1^2 ] {a^2\over m^2} + \O\left(\frac{a^3}{m^3}\right) \right\},
\end{equation} 
where we have
\begin{equation}
\sin\Theta_{analytic}= {3\sqrt{3m^2+a^2} \sqrt{r_*^2-2mr_*+a^2} \over r_*^2+3a^2},
\end{equation}
and
\begin{equation}
k_1 = {3\over2} \left[1-\left(\frac{3m}{r}\right)^2\right];
\qquad
k_2 = 
- {3\over 4} \left[1+\left(\frac{6m}{r}\right)^2-3\left(\frac{3m}{r}\right)^4\right].
\end{equation}

All of the angular dependence is hidden in $Q_1$ and $Q_2$:
\begin{equation}
Q_1 = -{2\over\sqrt{27} }\sin\Phi \sin\theta_* ,
\qquad
Q_2 = -{2\over27}(2+\sin^2\theta_* [1-2\sin^2\Phi]).
\end{equation}

The solid angle can then be calculated and is given as
\begin{eqnarray}
\label{Omega_2nd_order_Bb}
\Delta\Omega_* & =& 2\pi \left[1+{(r_*-3m)\sqrt{r_*^2+6mr_*-3a^2} \over r_*^2+3a^2}\right] 
\nonumber\\
&&
\qquad
 - {3\pi\over2} {\sqrt{1+6m/r_*-3a^2/r_*^2} \over 1+3a^2/r_*^2}{(1-2m/r_*)\over(1+6m/r_*)^2}{m^2\over r_*^2} 
 \nonumber\\
 &&\qquad\qquad\times
\left\{ 
\left[ -8 \hat k_1 +\sin^2\theta_*\hat k_2 \right] {a^2\over m^2} 
+ \O\left(\frac{a^4}{m^4}\right) \right\},
\end{eqnarray}
where now
\begin{equation}
\hat k_1 = \left[1+{3m\over r_*}\right]\left[1+{6m\over r_*}\right];
\end{equation}
and
\begin{equation}
\hat k_2 = 
\left[1+3\left(3m\over r_*\right)-2\left(3m\over r_*\right)^2-18\left(3m\over r_*\right)^3-12\left(3m\over r_*\right)^4\right].
\end{equation}

Overall, while in general very little can be done analytically, perturbatively one can calculate first and second order contributions to the escape cone and solid angle without the need of appealing to numerical methods. (While there is no physical or mathematical obstruction to going to third order in $a/m$, or even fourth order in $a/m$, the results are too messy to be useful.)

\section{Discussion}
So what have we learnt from this discussion? One important point is the intimate relationship between escape cone, capture cone, and silhouette. 
When phrased in terms of escape cones, the discussion above describes non-trivial near-horizon effects which physically describe the apparent dimming of light emitting objects as they approach the event horizon. These effects could, in theory, be observed as we increase our observational resolution power.
When phrased in terms of silhouettes, the discussion is particularly relevant to the ongoing observational programme of the Event Horizon Telescope, (which should more properly be referred to as the ``Near Horizon Telescope"). 

\enlargethispage{20pt}
Herein we have focussed on the low-rotation limit, extracting as much in the way of analytical insight as possible. We have  worked (in terms of the metric and other physically measurable quantities) to first and second order in $a/m$. This analysis is related to (but not identical to) the Lense--Thirring approach which (in terms of the tetrad) essentially works only to first order in $a/m$~\cite{Lense:1918, Mashhoon:1984,Baines:2020,Baines:2021a,Baines:2021b,Baines:2022}. 
The formulae we have developed, while intricate, give fully explicit and analytic control over the escape cones in the low-rotation limit.

\bigskip
\hrule\hrule\hrule

\section*{Acknowledgements}

JB was supported by  Victoria University of Wellington PhD Doctoral Scholarships.
\\
During early phases of this work MV was directly supported by the Marsden Fund, 
via a grant administered by the Royal Society of New Zealand.

\bigskip
\hrule\hrule\hrule
\addtocontents{toc}{\bigskip\hrule}
%
\setcounter{secnumdepth}{0}
\section[\hspace{14pt}  References]{}
%

\end{document}